\documentclass[useAMS,usenatbib]{mn2e}
\usepackage{times}
\usepackage{graphics,epsf}
\usepackage{amsmath}                % American Mathematical Society package
\usepackage{amsfonts}               % American Mathematical Society fonts
\usepackage{amssymb}                % American Mathematical Society symbol
\usepackage{epsfig}                 % EPS figures
\usepackage{rotating}
\usepackage{color}
\usepackage{tabularx}

\newcommand{\cm}{{~\rm cm}}
\newcommand{\s}{{~\rm s}}
\newcommand{\km}{{~\rm km}}
\newcommand{\g}{{~\rm g}}
\newcommand{\K}{{~\rm K}}

\newcommand{\yr}{{~\rm yr}}

\newcommand{\pc}{{~\rm pc}}

\newcommand{\AU}{{~\rm AU}}

\newcommand{\sr}{{~\rm sr}}
\def \apj{ApJ}
\def \aap{A\&A}

\def \mnras{MNRAS}
\def \apjl{ApJ Lett.}
\def \apjs{ApJ Suppl. Ser.}

\def \araa{ARA\&A}

\begin{document}

\title{The circumstellar matter of supernova 2014J and the core-degenerate scenario}

\author[Noam Soker] {Noam Soker\\ Department of Physics, Technion -- Israel Institute of Technology, Haifa 32000, Israel;\\soker@physics.technion.ac.il}
\maketitle

%%% \author{Noam Soker\altaffilmark{1}}
%%% \altaffiltext{1}{Department of Physics, Technion -- Israel Institute of Technology, Haifa 32000 Israel; soker@physics.technion.ac.il.}

\begin{abstract}
I show that the circumstellar matter (CSM) of the type Ia
supernova 2014J is too massive and its momentum too large to be
accounted for by any but the core-degenerate (CD) scenario for
type Ia supernovae. Assuming the absorbing gas is of CSM origin,
the several shells responsible of the absorption potassium lines
are accounted for by a mass loss episode from a massive asymptotic
giant branch star during a common envelope phase with a white
dwarf companion. The time-varying potassium lines can be accounted
for by ionization of neutral potassium and the Na-from-dust
absorption (NaDA) model. Before explosion some of the potassium
resides in the gas phase and some in dust. Weakening in absorption
strength is caused by potassium-ionizing radiation of the
supernova, while release of atomic potassium from dust increases
the absorption. I conclude that if the absorbing gas originated
from the progenitor of SN~2014J, then a common envelope phase took
place about 15,000 years ago, leading to the merging of the core
with the white dwarf companion, i.e., the core-degenerate
scenario. Else, the absorbing material is of interstellar medium
origin.
\end{abstract}


% ==========================================================
\section{INTRODUCTION}
\label{sec:introduction}
% ==========================================================

Type Ia supernovae (SN Ia) are thermonuclear explosions of white
dwarfs (WDs) accompanied by a complete destruction of the WD, or
at least one of the two interacting WDs (e.g.
\citealt{Maozetal2014}). There is as of yet no consensus on the
scenario that brings a WD or two to explode. In recent years there
have been several different scenarios, that were classified into
five categories by \cite{TsebrenkoSoker2015a}. We list them below
by their alphabetical order, and cite only a limited fraction of
the literature on each scenario.
 \newline
(a){\it The core-degenerate (CD) scenario} (e.g.,
\citealt{Sparks1974, Livio2003, KashiSoker2011, Soker2011,
IlkovSoker2012, IlkovSoker2013, Sokeretal2013}).
 \newline
(b){\it The double degenerate (DD) scenario} (e.g.,
\citealt{Webbink1984, Iben1984, vanKerkwijk2010, Loren2010,
Pakmoretal2013, Aznar2014}).
 \newline
(c){\it The double-detonation (DDet) mechanism} (e.g.,
\citealt{Woosley1994, Livne1995, Shenetal2013}.
 \newline
(d){\it The single degenerate (SD) scenario} (e.g.,
\citealt{Whelan1973, Nomoto1982, Han2004}).
 \newline
(e) \emph{The WD-WD collision (WWC) scenario} (e.g.,
\citealt{Raskinetal2009, Rosswogetal2009, Thompson2011,
KatzDong2012, Kushniretal2013}).

As there is no consensus on the SN Ia progenitor(s), it is crucial
to refer to all five scenarios (or categories of scenarios) when
confronting observations with SN Ia scenarios. In the present
study I examine some properties of the type Ia SN~2014J,
concentrating on the medium around it. I take into account limits
on the pre-explosion progenitor \citep{Kellyetal2014,
Nielsenetal2014, Lundqvistetal2015} and its mass loss
\citep{Perezetal2014}, and the properties of the absorption
potassium lines \citep{Ritcheyetal2014}, and their
time-variability \citep{Grahametal2015}.

>From radio observations \cite{Perezetal2014} limit the mass loss
rate from the progenitor of SN~2014J to $<7.0 \times 10^{-10}
M_\odot \yr^{-1}$ for a wind speed of $100 \km \s^{-1}$. This
contradicts many cases of the SD scenario, but not those with a
long delay to explosion \citep{DiStefanoetal2011, Justham2011}.
\cite{Perezetal2014} consider the DD scenario to be the
alternative to the SD scenario; without explanation they ignore
the other three scenarios listed above, and the presence of dense
circum-stellar matter (CSM) in some cases of the DD scenario
\citep{Moll+2014, Raskinetal2014, Levanonetal2015}. A similar but
weaker constraint is obtained from X-ray observations by
\cite{Marguttietal2014}, who also conduct a thorough discussion of
possible scenarios, leaving some room for the SD scenario with
time delay to explosion \citep{DiStefanoetal2011, Justham2011} and
for the CD scenario. One should note that the WWC scenario also
complies with these limits on the close CSM.
\cite{Grahametal2015}, on the other hand, attribute the
time-varying potassium lines from SN~2014J to a CSM, and argue
that their results tentatively support a SD scenario for SN2014J.

In this study I reexamine the claims for a CSM around SN 2014J,
and try to reconcile such a CSM with the CD scenario. I do not get
into the dispute whether the absorbing material is of CSM (e.g.,
\citealt{Foleyetal2014}) or of interstellar medium (ISM) origin
(e.g., \citealt{Johansson2014}), but rather discuss the
consequences of the case that the absorbing gas, or some of it, is
of CSM origin. In section \ref{sec:lines} I examine the potassium
lines, and in section \ref{sec:mass} I estimate the mass of the
absorbing gas based on the results of \cite{Grahametal2015}. Under
the assumption that the absorbing material, or part of it,
originated from the progenitor of SN~2014J I confront the
different scenarios with the CSM properties in section
\ref{sec:summary}.

% ==========================================================
\section{POTASSIUM ABSORPTION LINES}
\label{sec:lines}
% ==========================================================

The absorption potassium lines were studied by
\cite{Ritcheyetal2014} and \cite{Grahametal2015}. In examining the
higher-resolution spectrum presented by \cite{Grahametal2015} in
their figure 3, one can notice time variations of absorption lines
at three velocities. At velocities of $v=-144$, $-127\km \s^{-1}$,
the most blue shifted lines in K~I, the absorption lines were
weakening with time, while at the velocity of $-81 \km \s^{-1}$
the absorption line became stronger with time. No time variation
was found in any of the sodium lines.

The weakening of the potassium absorption lines can be attributed
to the ionization of neutral atoms by the SN radiation.
\cite{Grahametal2015} find that while K~I shows time-variation, no
such variation is observed in any of the Na~I line. For some Na~I
lines this invariability is due to their saturation.
\cite{Grahametal2015} argue that if the absorbing material resides
at $\approx 0.6-1.6 \times 10^{19} \cm$ from the SN, then the
potassium ionization fraction increases substantially with time,
while the sodium suffers only a moderate increase in ionization
fraction. This can account for the weakening of the $v=-144$ and
$-127\km \s^{-1}$ K~I absorption lines.

The observation that only the two most blue-shifted lines show
this behavior was attributed by \cite{Grahametal2015} to these
shells being the closest shells to he SN. The other shells in
their explanation have been slowed down by the ISM. As I show in
the next section, this explanation is problematic, as the implied
total mass in the shells in such a mass distribution is larger
than $8 M_\odot$. This is more than a progenitor of SN Ia can
supply.

\cite{Grahametal2015} do not study the strengthening of the $-81
\km \s^{-1}$ K~I absorption line. The strengthening of absorption
lines can be attributed to either the opposite process of
recombination, or to the release of more neutral atoms from dust.
The process by which the SN Ia radiation releases sodium from dust
residing at $\approx 1 \pc$ from the SN, the Na-from-dust
absorption (NaDA) model \citep{Soker2014}, is similar to the
processes by which solar radiation releases sodium from cometary
dust when comets approach a distance of $\la 1 \AU $ from the Sun.
The NaDA model was suggested to account for time-varying sodium
lines in SN Ia (for observations of Na~I lines in SN Ia see, e.g.,
\citealt{Patatetal2007, Simonetal2009, Sternbergetal2011,
Sternbergetal2014, Maguireetal2013}). Dust in the solar system is
known to release potassium in addition to sodium, e.g.,
\cite{Fulleetal2013}. I here suggest that the deepening of the
$-81 \km \s^{-1}$ absorption line in the spectrum of SN 2014J is a
result of potassium release from dust, as in the NaDA model.

The velocity of the line is explained within the CD scenario, if
the absorbing gas is indeed a CSM and not an ISM, in the following
way. Some or all of the absorbing clouds/shells were ejected by
the progenitor within a very short time period, the common
envelope (CE) phase of the progenitor binary sytem. The shells are
actually clumpy nebula expanding in a linear relation between
distance and velocity. Namely, the $-81 \km \s^{-1}$ gas is at a
distance of $81/144$ times the distance of the $-144 \km \s^{-1}$
gas from the explosion. Such expanding clumpy nebulae with a
linear distance-velocity relation and expansion velocities of up
to few hundreds $\km \s^{-1}$ is observed in some planetary
nebulae (PNe), e.g., NGC~6302 (e.g., \citealt{Meaburnetal2008,
Szyszkaetal2011}) and the pre-PN M1-92 (IRAS 19343+2926;
\citealt{Bujarrabaletal1998}).

 My suggestion that the absorbing gas of the most blue-shifted
lines lies at the largest distances from the explosion is opposite
to the claim of \cite{Grahametal2015}, but gives a  lower CSM mass
that can be compatible with a CSM, as is shown in the next
section. The proposed scenario implies that SN~2014J is a
SN-Inside a PN (SNIP; \citealt{DickelJones1985,
TsebrenkoSoker2013, TsebrenkoSoker2015a, TsebrenkoSoker2015b}).

% ==========================================================
\section{THE MASS OF THE ABSORBING GAS}
\label{sec:mass}
% ==========================================================

I first consider the three potassium absorption lines where
time-variability is observed \citep{Grahametal2015}.  The column
densities of the neutral potassium of the gas absorbing in the
$-144 \km \s^{-1}$, $-127 \km \s^{-1}$, and $-81 \km \s^{-1}$
lines are $0.50 \times 10^{11}\cm^{-2}$, $4.0 \times 10^{11}
\cm^{-2}$, and $1.1 \times 10^{11} \cm^{-2}$, respectively
\citep{Ritcheyetal2014}. \cite{Grahametal2015} find a somewhat
higher column density of $0.80 \times 10^{11}\cm^{-2}$ for the
$-144 \km \s^{-1}$ shell. Using higher column densities will
strengthen the conclusions reached here.

The mass of a shell, or a partial shell, at a radius $r_s$ with am
atomic potassium column density of $N({\rm K I})$ and a solar
composition \citep{Asplundetal2009} is given by
\begin{equation}
M_s= 0.53
 \left( \frac{N({\rm K I})}{10^{11} \cm^{-2} } \right)
\left( \frac{r_s}{2 \pc} \right)^{2}
 \frac {\beta}{\xi}   M_\odot,
  \label{eq:shellmass}
\end{equation}
where $4 \pi {\beta}$ is the solid angle covered by the shell and
${\xi^{-1}}$ is the fraction of the potassium in the atomic phase.
Both $\beta<1$ and $\xi$ are likely to be less than unity.

The scaling of $r_s=2 \pc$ is the minimum radius allows by
\cite{Grahametal2015} in their explanation for the time variation
of the two most blue-shifted K~I lines. For the values of $r_s=2
\pc$ and ${\beta}/{\xi} =1$, the masses of the two shells are
$M(-144)=0.26 M_\odot$ and $M(-127)=2.1 M_\odot$. If the mass is
spread over a radial distance of $\approx 1 \pc$, the duration of
the mass loss process is $\approx 1 \pc / 140 \km \s^{-1} \approx
7000 \yr$. The mass loss rate is $\approx 2 M_\odot /7000 \yr = 3
\times 10^{-4} M_\odot \yr^{-1}$. This is a very high mass loss
rate, more than is expected from any model for the SD scenario
where a low mass transfer rate is required from a giant to the WD
companion. The existence of shells suggests that the mass loss
rate was over a shorter time scale than $7000 \yr$, and mass loss
rate was higher even.

As mentioned in section \ref{sec:lines}, the existence of clumps
with a velocity spread is observed in some PNe and pre-PNe, and in
some cases with a linear velocity-distance relation. In these
cases the mass ejection was a short event, most probably a CE
phase, and possibly an intermediate-luminosity optical transients
(ILOT) event \citep{KashiSoker2011, AkashiSoker2013}. A bipolar PN
is formed with a solid angle coverage of $\beta<1$.

To examine this CD scenario, I do the following exercise. I take
the blue-shifted shells listed by \cite{Ritcheyetal2014} to
originate from the progenitor of SN~2014J, and obey
\begin{equation}
 r_s(i)=t_{\rm ejc} v_s(i) = 1.53
\left( \frac{t_{\rm ejc}}{1.5 \times 10^4 \yr} \right) \left(
\frac{v_s(i)}{100 \km \s^{-1} } \right) \pc,
  \label{eq:rs}
\end{equation}
where $r_s(i)$ and $v_s(i)$ are the radius and the velocity of
shell $i$, and $t_{\rm ejc}$ is the time since the shells were
ejected in a short event by the progenitor of SN~2014J. Equation
(\ref{eq:rs}) takes SN~2014J to be in the rest  frame of M82.
However, SN~2014J is in the approaching side of M82, and it is
likely that the relative shells velocities to SN~2014J are
somewhat lower. In that case the absorbing gas likelihood to be of
ISM origin is higher. In the present study I examine the
implications of the absorbing gas being of CSM origin, and take
the shells distances from the explosion according to equation
(\ref{eq:rs}).

The outer shell has $v=-144 \km \s^{-1}$ with $r_s(-144)=2.2
(t_{\rm ejc}/1.5 \times 10^4 \yr) \pc$. After summing over all the
shells I find the CSM mass in the CD scenario (assuming all
blue-shifted shells belong to the CSM) to be
\begin{equation}
M_{\rm CSM-Cd}= 5.4
 \left( \frac{t_{\rm ejc}}{1.5 \times 10^4 \yr} \right)
 \frac{\bar \beta}{\bar \xi}
  M_\odot,
  \label{eq:csmt}
\end{equation}
where $\bar {\beta}$ and $\bar \xi$ are average values for the
solid angle covered by the shells and the fraction of potassium in
the atomic state; both are expected to be less that unity. The
total mass of the three shells that show time-variability with the
same scaling is $2.2 M_\odot$. Over all, the progenitor had to
expel $\approx 2-6 M_\odot$ over a relatively short time of $\la
10^4 \yr$. The mass loss rate during the formation of the shells,
if they are indeed CSM, was $\gg 10^{-4} M_\odot \yr$. A single
star of mass $<8 M_\odot$ is not expected to have such a high mass
loss rate. A CE interaction seems the most natural explanation for
such a mass loss rate.

Other relevant points to the proposed mass distribution are as
follows.
 \newline
 (1) If all shells reside outside $2 \pc$, as suggested by \cite{Grahametal2015}, then the mass
with the above scaling is $> 8M_\odot$. Larger than what a
progenitor of a SN I can supply.
 \newline
  (2) \cite{Johansson2014} limit the dust mass within $\approx 10^{17} \cm$
from SN~2014J to be $\la 10^{-5} M_\odot$. The inner shell in the
structure discussed above is taken to be the one with a velocity
of $ - 19 \km \s^{-1}$. Its distance from the center is $r_s(-19)=
9 \times 10^{17} (t_{\rm ejc}/1.5 \times 10^4 \yr) \cm$, and its
mass is $M_s(-19)=0.004 {\beta}/{\xi} M_\odot$ under the
assumptions employed here. The location and mass of the shell is
compatible with the limit set by \cite{Johansson2014}. If the mass
distribution employed here is correct, then within about two years
from the explosion, sometime in 2016, the infrared radiation from
warming CSM dust might start to increase.
 \newline
 (3) \cite{Nielsenetal2014} argue that SN~2014J comes
from a young population. A mass transfer from the initially more
massive primary star, $M_{10} \approx 5-8 M_\odot$, to the
secondary could leave a massive secondary on the main sequence,
$M_{2}\approx 7-8 M_\odot$, that expelled such a large mass during
the CE phase \citep{IlkovSoker2013}, as in the case of PTF~11kx
\citep{Sokeretal2013}. (Note that in PTF~11kx the CSM was much
closer to the exploding star.)
 \newline
(4) In this study I do not deal with the question of why three
shells show time-variability, but not the others. It might be
related to the density of the shells, or some other properties.
This might actually be an argument for that the shells are ISM
shells, and not CSM shells. However, then one will have to explain
why most of the shells show blue-shifted absorption. I do note
that even if only the three time-variable shells belong to the
CSM, the discussion to come still holds.

% ==========================================================
\section{DISCUSSION AND SUMMARY}
 \label{sec:summary}
% ==========================================================

In the present study I do not get into the question whether the
gas responsible for the potassium absorption lines is of CSM or
ISM origin. This question is not settled yet. I consider the
implications of the CSM case. Namely, where some, or all, of the
absorbing clouds/shells were originated from the progenitor of
SN~2014J, as argued most recently by \cite{Grahametal2015}. But
one must keep in mind that it is quite possible that all the
absorbing gas clouds/shells are ISM. In any case, the discussion
in this section is relevant for future observations of CSM around
SN Ia, as I confront each of the five scenarios listed in section
\ref{sec:introduction} with the properties of the absorbing gas of
SN~2014J if it is of CSM origin.

% =================
\subsection{The core-degenerate (CD) scenario}
 \label{sec:CD}
% =================

In the CD scenario the pre-explosion mass ejection episode in
SN~2014J lasted for a relatively short time, at most few hundreds
of years, which is the CE evolution time. The distances of the
shells from the explosion is more or less their velocity times the
time from mass ejection to explosion, as given in equation
(\ref{eq:rs}).

The merger of the core with the WD could have energize the shells
to the high velocities, up to $140 \km \s^{-1}$, more than the
typical velocities of post-CE nebulae, few$\times 10 \km \s^{-1}$.
Taking the shells radii used in previous sections (based on
\citealt{Grahametal2015}), the age of the nebula is $\approx 1.5
\times 10^4 \yr$. The evolution of cores of AGB stars with masses
of $\ga 0.6 M_\odot$ is rapid enough that their luminosity and
temperature after $\approx 1.5 \times 10^4 \yr$ of post-AGB
evolution are $L_c \approx 200-250 L_\odot$ and $T_{\rm eff}
\approx 10^5 \K$, respectively, e.g., \cite{Bloecker1995}. This
luminosity is below the upper limit on the luminosity of the
progenitor of SN~2014J of $\approx 10^3 L_\odot$
\citep{Kellyetal2014}.

The ejection of the entire envelope explains the large mass of the
shells, $\approx 2-6 M_\odot$, found in section \ref{sec:mass}.

By the post-CE time of $1.5 \times 10^4 \yr$ the wind mass loss
rate from the central star is $\approx 10^{-10} M_\odot \yr^{-1}$
and its velocity is thousands of $\km \s^{-1}$
\citep{Schonberneretal2014}. This gives a very low density gas
close to the explosion, $r \la 10^{18} \cm$, that is safely
compatible with the limit set by \cite{Perezetal2014} and
\cite{Marguttietal2014} on the mass loss rate from the progenitor
hundreds of year before explosion.

Despite the hot-luminous central star, most of the material can be
stay neutral, as required by the observed absorption. The hydrogen
ionizing photon rate per unit solid angle from a post-AGB star
with an effective temperature of $T_{\rm eff}=10^5 \K$ is
\begin{equation}
\dot n_\gamma ({10^5K})= 1.7 \times 10^{45} \left( \frac {L_c}{300
L_\odot} \right) \s^{-1} \sr^{-1}.
  \label{eq:ionize}
\end{equation}
The hydrogen recombination rate of a partial shell covering a
solid angle of $4 \pi \beta$, for a gas temperature of $10^4 \K$,
and almost fully ionized is
\begin{equation}
\begin{split}
\dot n_{\rm rec-i} = & 1.9 \times 10^{45} \beta^{-1}
  \left(\frac{m_i}{0.3 M_\odot} \right)^2 \\
 & \times
 \left( \frac{r_i}{10^{18} \cm}
\right)^{-3} \left( \frac{\Delta r_i}{0.1 r_i} \right)^{-1}
\s^{-1} \sr^{-1}.
 \label{eq:recom}
 \end{split}
\end{equation}
where $r_i$ is the radius, $\Delta _i$ is the thickness, and $m_i$
is the mass  of the ionized shell (or partial shell).

Comparing equations ( \ref{eq:ionize})  and ( \ref{eq:recom})
shows that an inner shell of about $10 \%$ of the total shell
mass,  i.e., $m_i \approx 0.2-0.6 M_\odot$, can shield the outer
shells from the pre-explosion ionizing radiation. This can account
for the neutral potassium and sodium phase that is required for
the observed absorption.

 %% USE: http://www.spectralcalc.com/blackbody_calculator/blackbody.php

% =================
\subsection{The double-degenerate (DD) scenario}
 \label{sec:DD}
% =================

The massive shells require a CE evolution. The shells were ejected
$<2 \times 10^4 \yr$ ago, which is the post-CE period. This
implies that (a) the core is still very hot during the WD-core
merger, and (b) that for about a thousand years the core is still
larger than a cold WD, a radius of $\ga 0.1 R_\odot$
\citep{Bloecker1995}. For gravitational radiation to bring the
core-WD system to merge within $<2 \times 10^4 \yr$ the orbital
separation should be $ \la 0.1 R_\odot$. At this separation tidal
interaction with the larger core would cause the merger rather
than gravitational radiation. For these two reasons the CD
scenario is a more appropriate description of the evolution than a
DD scenario.

% =================
\subsection{The double-detonation (DDet) scenario}
 \label{sec:DDet}
% =================

There are two issues here. (1) The time from the formation of a
donor He-WD to the mass transfer phase form the He-WD to the CO-WD
in the DDet scenario is very long \citep{Shenetal2013}, much
longer than the post-CE evolution of SN~2014J. (2) The expected
CSM mass in the DDet scenario is very low, $\ll 1 M_\odot$
\citep{Shenetal2013}. For these two reason the DDet scenario
cannot account for the potassium absorbing gas of SN~2014J if it
is of CSM origin. {{{ \cite{Lundqvistetal2015} put further
constraint on any helium donor companion to SN~2014J to be at a
relatively large separation. }}}

% =================
\subsection{The single-degenerate (SD) scenario}
 \label{sec:SD}
% =================

In the SD scenario there is no strong interaction between the
donor star and the WD. The mass loss rate is not expected to be as
high as inferred in section \ref{sec:mass}, $\ga 3 \times 10^{-4}
M_\odot \yr^{-1}$. The radial momentum discharge of the shells is
defined as the radial momentum of the shells divided by the
ejection time. I find this value for the shells, if of CSM origin,
to be $\dot p \ga 10^{29} \g \cm \s^{-2}$. The radiation momentum
discharge is $L_\ast/c= 1.3\times 10^{28} (L_\ast/10^5 L_\odot) \g
\cm \s^{-2}$. A wind from a single giant star of mass $<8 M_\odot$
cannot explain the momentum in the absorbing shells.

Another problem is the time span between mass ejection and
explosion. Since the present luminosity of the companion is
limited to $\la 10^3 L_\odot$ \citep{Kellyetal2014}, any companion
with such a massive CSM must be brighter on the asymptotic giant
branch (AGB). To explain the massive CSM, the giant donor should
have lost most of its envelope within $<2 \times 10^4 \yr$ prior
to explosion. In the SD scenario there is no explanation why
explosion should occur within this time scale. This is unlike the
CD scenario, where the time scale is explained as a post-CE
evolution.

% =================
\subsection{The WD-WD collision (WWC) scenario}
 \label{sec:WWC}
% =================

In the WWC scenario the two progenitors of the two WDs should have
no interaction at all before the final collision. As explained for
the SD scenario above, a single AGB star cannot account for the
momentum and mass loss rate that formed the shells, if of CSM
origin.

This adds to the several problems the WWC scenario encounters when
compared with observations \citep{TsebrenkoSoker2015a}, despite
the one strong character that the ignition is easy to achieve in
the collision process. The most sever problem of the WWC scenario
is that it cannot account for more than about one percent of all
SN Ia (e.g., \citealt{Sokeretal2014}; see section 2 in the first
astro-ph version of that paper, arXiv:1309.0368v1).

Manganese production is not accounted for in the WWC scenario.
\cite{Tamagawaetal2009} identified X-ray lines from manganese in
the Tycho SNR. \cite{Seitenzahletal2013} further claim that at
least $50\%$ of all SN Ia come from near Chandrasekhar mass
($M_{\rm Ch}$) WDs, as the density required to synthesize
manganese is $\rho \ga 2 \times 10^8 \g \cm^{-1}$
(\citealt{Seitenzahletal2013} and references therein). The WWC
scenario does not reach these densities, and cannot account for
the production of manganese. The same problem holds for the DDet
scenario.

Another problem seems to be the iron distribution in the SN
remnant (SNR). \cite{Dongetal2014} consider the presence of
doubly-peaked line profiles of cobalt and iron as a possible
smoking gun for the WWC scenario.  Although \cite{Dongetal2014}
account for the doubly-peaked lines, it seems that their predicted
iron distribution is in contradiction with iron distribution of
two resolved SNRs. \cite{Yamaguchietal2012} present the 2D iron
distribution in the SNR of the possible SN~Ia~G344.7-0.1, and
\cite{Fesenetal2015} present the iron and calcium 2D distributions
in the SNR~1885 in the Andromeda galaxy. In both SNRs the iron
distribution is clumpy, and does not show two prominent
distributions as expected from the WWC scenario.

% =================
\subsection{Summary}
 \label{subsec:summary}
% =================

Under the assumption that the gas responsible of the potassium
absorption lines in SN~2014J, either only time-variable lines or
all lines, is CSM formed by the progenitor of SN~2014J, I found in
this study that the CD scenario is the only scenario that can
account for the mass and momentum of the absorbing gas.

This does not mean that the issue is completely settled. There are
several open questions in the CD scenario that still need to be
worked out. Specifically for SN~2014J the time-variability of
three absorption lines and invariability of the other lines should
be explained. More generally, the processes of the core-WD merger,
the evolution till ignition, and the ignition must be worked out
in the CD scenario.

If the absorbing material in SN~2014J is indeed CSM, it adds to
PTF~11kx in supporting the CD scenario. In PTF~11kx the CD
scenario is the only one among the scenarios studied here that can
account for the mass and properties of the CSM
\citep{Sokeretal2013}. Including other observational properties of
SN Ia that were listed by \cite{TsebrenkoSoker2015a} in their
table 1, I find the CD scenario to be the favorable one for SN Ia,
possibly together with the DD scenario. The automatic attribution
of any CSM material around SN Ia to the SD scenario should be
abandon. The CD scenario does much better.

This research was supported by the Asher Fund for Space Research
at the Technion.

%--------------------------
%\footnotesize
{}


\begin{thebibliography}{}

\bibitem[Akashi \& Soker(2013)]{AkashiSoker2013} Akashi, M., \& Soker, N.\
2013, \mnras, 436, 1961

\bibitem[Asplund et al.(2009)]{Asplundetal2009} Asplund, M., Grevesse, N., Sauval,
A.~J., \& Scott, P.\ 2009, \araa, 47, 481

\bibitem[Aznar-Sigu{\'a}n et al.(2013)]{Aznar2014} Aznar-Sigu{\'a}n, G., Garc{\'{\i}}a-Berro, E., Lor{\'e}n-Aguilar, P., Jos{\'e}, J., \& Isern, J.\ 2013, \mnras, 434, 2539

\bibitem[Bloecker(1995)]{Bloecker1995} Bloecker, T.\ 1995, \aap, 299, 755

\bibitem[Bujarrabal et al.(1998)]{Bujarrabaletal1998} Bujarrabal, V., Alcolea, J.,
Sahai, R., Zamorano, J., \& Zijlstra, A.~A.\ 1998, \aap, 331, 361

\bibitem[Dickel \& Jones(1985)]{DickelJones1985} {{{  Dickel, J.~R., \& Jones, E.~M.\ 1985, \apj,
288, 707 }}}

\bibitem[Di Stefano et al.(2011)]{DiStefanoetal2011}  Di Stefano, R.,
Voss, R., \& Claeys, J.~S.~W.\ 2011, \apjl, 738, LL1

\bibitem[Dong et al.(2014)]{Dongetal2014} Dong, S., Katz, B., Kushnir, D., \& Prieto, J.~L.\ 2014, arXiv:1401.3347

\bibitem[Fesen et al.(2015)]{Fesenetal2015} Fesen, R.~A., Hoeflich,
P., \& Hamilton, A.~J.~S.\ 2015, arXiv:1412.3815

\bibitem[Foley et al.(2014)]{Foleyetal2014} Foley, R.~J., Fox, O.~D.,
McCully, C., et al.\ 2014, \mnras, 443, 2887

\bibitem[Fulle et al.(2013)]{Fulleetal2013} Fulle, M., Molaro, P.,
Buzzi, L., \& Valisa, P.\ 2013, \apjl, 771, LL21

\bibitem[Graham et al.(2015)]{Grahametal2015} Graham, M.~L., Valenti,
S., Fulton, B.~J., et al.\ 2015, arXiv:1412.0653

\bibitem[Han \& Podsiadlowski(2004)]{Han2004} Han, Z., \& Podsiadlowski, P.\ 2004, \mnras, 350, 1301

\bibitem[Iben \&  Tutukov(1984)]{Iben1984} Iben, I.,  Jr., \& Tutukov, A.~V.\ 1984, \apjs, 54, 335

\bibitem[Ilkov \& Soker(2012)]{IlkovSoker2012} Ilkov, M., \& Soker, N.\ 2012, \mnras, 419, 1695

\bibitem[Ilkov \& Soker(2013)]{IlkovSoker2013} Ilkov, M., \& Soker, N.\ 2013, \mnras, 428, 579

\bibitem[Johansson et al.(2014)]{Johansson2014} Johansson, J.,
Goobar, A., Kasliwal, M.~M., et al.\ 2014, arXiv:1411.3332

\bibitem[Justham(2011)]{Justham2011} Justham, S.\ 2011, \apjl, 730, LL34

\bibitem[Kashi \& Soker(2011)]{KashiSoker2011} Kashi, A., \& Soker, N.\ 2011, \mnras, 417, 1466

\bibitem[Katz \& Dong(2012)]{KatzDong2012} Katz, B., \& Dong, S.\ 2012, arXiv:1211.4584

\bibitem[Kelly et al.(2014)]{Kellyetal2014} Kelly, P.~L., Fox, O.~D.,
Filippenko, A.~V., et al.\ 2014, \apj, 790, 3

\bibitem[Kushnir et al.(2013)]{Kushniretal2013} Kushnir, D., Katz, B., Dong, S., Livne, E., \& Fern{\'a}ndez, R.\ 2013, \apjl, 778, L37

\bibitem[Lundqvist et al.(2015)]{Lundqvistetal2015} {{{ Lundqvist, P.,
Nyholm, A., Taddia, F., et al.\ 2015, arXiv:1502.00589 }}}

\bibitem[Levanon et al.(2015)]{Levanonetal2015} Levanon, N., Soker, N.,
\& Garc{\'{\i}}a-Berro, E.\ 2015, \mnras, 447, 2803

\bibitem[Livio \& Riess(2003)]{Livio2003} Livio, M., \& Riess, A.~G.\ 2003, \apjl, 594, L93

\bibitem[Livne  \&  Arnett(1995)]{Livne1995}  Livne,  E.,  \&  Arnett,  D.\ 1995, \apj, 452, 62

\bibitem[Lor{\'e}n-Aguilar et al.(2010)]{Loren2010} Lor{\'e}n-Aguilar, P., Isern, J., \& Garc{\'{\i}}a-Berro, E.\ 2010, \mnras, 406, 2749

\bibitem[Maguire et al.(2013)]{Maguireetal2013} Maguire, K., Sullivan,
M., Patat, F., et al.\ 2013, \mnras, 436, 222

\bibitem[Maoz et al.(2014)]{Maozetal2014} Maoz, D., Mannucci, F., \& Nelemans, G.\ 2014, arXiv:1312.0628

\bibitem[Margutti et al.(2014)]{Marguttietal2014} Margutti, R.,
Parrent, J., Kamble, A., Soderberg, A. M., Foley, R. J.,
Milisavljevic, D., Drout, M. R., \& Kirshner, R.\ 2014, \apj, 790,
52

\bibitem[Meaburn et al.(2008)]{Meaburnetal2008} Meaburn, J., Lloyd, M.,
Vaytet, N.~M.~H., \& L{\'o}pez, J.~A.\ 2008, \mnras, 385, 269

\bibitem[Moll et al.(2014)]{Moll+2014} Moll, R., Raskin, C., Kasen, D., \& Woosley, S.~E.\ 2014, \apj, 785, 105

\bibitem[Nielsen et al.(2014)]{Nielsenetal2014} Nielsen, M.~T.~B.,
Gilfanov, M., Bogd{\'a}n, {\'A}., Woods, T.~E., \& Nelemans, G.\
2014, \mnras, 442, 3400

\bibitem[Nomoto(1982)]{Nomoto1982} Nomoto, K.\ 1982, \apj, 253, 798

\bibitem[Pakmor et al.(2013)]{Pakmoretal2013} Pakmor, R., Kromer, M.,
Taubenberger, S., \& Springel, V.\ 2013, \apjl, 770, L8

\bibitem[Patat et al.(2007)]{Patatetal2007} Patat, F., Chandra, P., Chevalier, R., et al.\ 2007, Science, 317, 924

\bibitem[P{\'e}rez-Torres et al.(2014)]{Perezetal2014}
P{\'e}rez-Torres, M.~A., Lundqvist, P., Beswick, R.~J., et al.\
2014, \apj, 792, 38

\bibitem[Raskin et al.(2014)]{Raskinetal2014} Raskin, C., Kasen, D.,
Moll, R., Schwab, J., \& Woosley, S.\ 2014, \apj, 788, 75

\bibitem[Raskin et al.(2009)]{Raskinetal2009} Raskin, C., Timmes,
F.~X., Scannapieco, E., Diehl, S., \& Fryer, C.\ 2009, \mnras,
399, L156

\bibitem[Ritchey et al.(2014)]{Ritcheyetal2014} Ritchey, A.~M., Welty,
D.~E., Dahlstrom, J.~A., \& York, D.~G.\ 2014, arXiv:1407.5723

\bibitem[Rosswog et al.(2009)]{Rosswogetal2009} Rosswog, S., Kasen, D.,
Guillochon, J., \& Ramirez-Ruiz, E.\ 2009, \apjl, 705, L128

\bibitem[Sch{\"o}nberner et al.(2014)]{Schonberneretal2014}
Sch{\"o}nberner, D., Jacob, R., Lehmann, H., Hildebrandt, G.,
Steffen, M., Zwanzig, A., Sandin, C., \& Corradi, R. L. M.\ 2014,
Astronomische Nachrichten, 335, 378

\bibitem[Seitenzahl et al.(2013)]{Seitenzahletal2013}
 Seitenzahl, I.~R., Cescutti, G., R{\"o}pke, F.~K., Ruiter, A.~J., \&
 Pakmor, R.\ 2013, \aap, 559, L5

\bibitem[Shen et al.(2013)]{Shenetal2013} Shen, K.~J., Guillochon, J., \& Foley, R.~J.\ 2013, \apjl, 770, L35

\bibitem[Simon et al.(2009)]{Simonetal2009} Simon, J.~D., Gal-Yam, A., Gnat, O., et al.\ 2009, \apj, 702, 1157

\bibitem[Soker(2011)]{Soker2011} Soker, N.\ 2011, arXiv:1109.4652

\bibitem[Soker(2014)]{Soker2014} Soker, N.\ 2014, \mnras, 444, L73

\bibitem[Soker et al.(2014)]{Sokeretal2014} Soker, N., Garc{\'{\i}}a-Berro, E., \& Althaus, L.~G.\ 2014, \mnras, 437, L66

\bibitem[Soker et al.(2013)]{Sokeretal2013} Soker, N., Kashi, A., Garc\'ia-Berro E., Torres, S., \& Camacho, J.\ 2013, \mnras, 431, 1541

\bibitem[Sparks \& Stecher(1974)]{Sparks1974}  Sparks, W.~M., \&  Stecher, T.~P.\ 1974, \apj, 188, 149

\bibitem[Sternberg et al.(2011)]{Sternbergetal2011} Sternberg, A., Gal-Yam, A., Simon, J.~D., et al.\ 2011, Science, 333, 856

\bibitem[Sternberg et al.(2014)]{Sternbergetal2014} Sternberg, A., Gal Yam, A., Simon, J.~D., et al.\ 2014, \mnras, 443, 1849

\bibitem[Szyszka et al.(2011)]{Szyszkaetal2011} Szyszka, C., Zijlstra,
A.~A., \& Walsh, J.~R.\ 2011, \mnras, 416, 715

\bibitem[Tamagawa et al.(2009)]{Tamagawaetal2009} Tamagawa, T., Hayato, A., Nakamura, S., et al.\ 2009, PASJ, 61, 167

\bibitem[Thompson(2011)]{Thompson2011} Thompson, T.~A.\ 2011, \apj, 741, 82

\bibitem[Tsebrenko \& Soker(2013)]{TsebrenkoSoker2013} Tsebrenko, D., \& Soker, N.\ 2013, \mnras, 435, 320

\bibitem[Tsebrenko \& Soker(2015a)]{TsebrenkoSoker2015a} Tsebrenko, D., \& Soker, N.\ 2015a, \mnras, 447, 2568

\bibitem[Tsebrenko \& Soker(2015b)]{TsebrenkoSoker2015b} Tsebrenko, D., \& Soker, N.\ 2015b,  arXiv:1407.6231

\bibitem[van Kerkwijk et al.(2010)]{vanKerkwijk2010} van Kerkwijk, M.~H., Chang, P., \& Justham, S.\ 2010, \apjl, 722, L157

\bibitem[Webbink(1984)]{Webbink1984} Webbink,  R.~F.\ 1984, \apj, 277, 355

\bibitem[Whelan \& Iben(1973)]{Whelan1973} Whelan, J., \& Iben, I., Jr.\ 1973, \apj, 186, 1007

\bibitem[Woosley  \&  Weaver(1994)]{Woosley1994}  Woosley,  S.~E., \& Weaver, T.~A.\ 1994, \apj, 423, 371

\bibitem[Yamaguchi et al.(2012)]{Yamaguchietal2012} Yamaguchi, H.,
Tanaka, M., Maeda, K., Slane, P. O., Foster, A., Smith, R. K.,
Katsuda, S., \& Yoshii, R.\ 2012, \apj, 749, 137

\end{thebibliography}
\end{document}